# Single and Double Hole Quantum Dots in Strained Ge/SiGe Quantum Wells

Will J. Hardy[a]*, C. T. Harris[a], Y.-H. Su[b], Y. Chuang[b], J. Moussa[a], Leon Maurer[a], Jiun-Yun Li[b], Tzu-Ming Lu[a], and Dwight R. Luhman[a].

[a]Sandia National Laboratories, Albuquerque, NM 87123, United States

[b]Department of Electrical Engineering, National Taiwan University, Taipei, Taiwan

ABSTRACT

Even as today's most prominent spin-based qubit technologies are maturing in terms of capability and sophistication, there is growing interest in exploring alternate material platforms that may provide advantages, such as enhanced qubit control, longer coherence times, and improved extensibility. Recent advances in heterostructure material growth have opened new possibilities for employing hole spins in semiconductors for qubit applications. Undoped, strained Ge/SiGe quantum wells are promising candidate hosts for hole spin-based qubits due to their low disorder, large intrinsic spin-orbit coupling strength, and absence of valley states. Here, we use a simple one-layer gated device structure to demonstrate both a single quantum dot as well as coupling between two adjacent quantum dots. The hole effective mass in these undoped structures, $m^* \sim 0.08\,m_0$, is significantly lower than for electrons in Si/SiGe, pointing to the possibility of enhanced tunnel couplings in quantum dots and favorable qubit-qubit interactions in an industry-compatible semiconductor platform.

INTRODUCTION

Recent years have yielded considerable progress in semiconductor quantum dot platforms and spin-based qubit demonstrations.[1–9] Silicon has emerged as a leading contender for hosting spin-based qubits due to long coherence times, made possible through isotopic enrichment, and the potential for complementary metal-oxide-semiconductor (CMOS) processing. Si/SiGe heterostructures have received considerable attention due to the low-disorder interface seen by the electron spin, leading to many important results (see, *e.g.*, the SiGe results from the above reference list). Despite these achievements, progress has been challenging due in part to material-related issues, such as inconsistent separation between the nearly degenerate valley states in the Si/SiGe system.[10–14]

A compelling alternative approach for hosting semiconductor spin-based qubits, which could potentially circumvent these challenges without sacrificing the clean environment of Si/SiGe, is the germanium-rich heterostructure system, Ge/SiGe. The material stack features a pure strained Ge quantum well with a SiGe barrier layer on either side containing majority Ge. The dominant charge carriers in Ge/SiGe are holes (rather than electrons in traditional Si/SiGe), with the heavy-hole and light-hole bands split by up to 100 meV as a result of confinement and strain.[15] The ground state heavy-hole band does not include nearly degenerate states, in contrast to the valley states in Si.[16] An additional benefit of hole-based qubits is that the *p*-orbital character of holes suggests that their wavefunction vanishes at the nucleus,[17–20] suppressing qubit decoherence via hyperfine interactions. In addition, both Si and Ge contain abundant naturally occurring spinless isotopes allowing for further suppression of hyperfine-induced decoherence effects through isotopic enrichment.

The relatively large cubic Rashba spin-orbit coupling (SOC) strength found in planar Ge/SiGe[15,21] provides an intrinsic mechanism for all-electrical qubit control via electric dipole spin resonance (EDSR).[17] The ability to manipulate qubit states by applying microwave electric fields directly to a quantum dot confinement electrode provides the possibility for simplified device layout schemes, in contrast to single-spin electron systems in silicon that have used on-chip micromagnets or microwave striplines for qubit control. Further, the large out-of-plane g-factor of up to g ~ 28 in the two-dimensional Ge/SiGe system[22,23] means that a relatively small externally applied magnetic field can be used to set the rotation frequency and facilitate readout in a single spin-based qubit.

A very recent demonstration[24] of a single quantum dot in planar Ge/SiGe showed that superconductivity can be induced in the quantum well via proximal aluminum leads, with the induced superconductivity extending over length scales of several microns,[25] which could enable hybrid quantum dot-superconductor devices. Other work[26–30] has made use of quantum dots formed in 'hut wires,' which are Ge nanowires of triangular cross section, grown by molecular beam epitaxy, in which single spin readout and EDSR have been reported.[31] A drawback to the latter geometry is that the wire growth locations are non-deterministic and non-planar, which complicates device fabrication.

Here, we demonstrate the feasibility of making both single and double quantum dots in a planar Ge/SiGe quantum well platform using a simple device layout, establishing the first steps toward a hole spin based qubit in this system. We demonstrate lithographic quantum dots using a single metal layer gate geometry.[32] Coulomb blockade oscillations are observed, along with Coulomb diamonds that show increasing charging energy as a function of gate bias. Coupling between two adjacent quantum dots is also demonstrated through the observation of a honeycomb pattern in a charge stability diagram. Additional transport measurements yield a value of the hole effective mass of $m^* \sim 0.08\ m_0$, significantly lower than for electrons in Si/SiGe. These results suggest that the Ge/SiGe system has excellent promise for novel spin-based qubit applications.

RESULTS AND DISCUSSION

The quantum dot devices featured in the following sections, as well as Hall bars used for initial material characterization, were prepared from two similar wafers grown back-to-back. The undoped, strained Ge/SiGe heterostructures were grown on Si substrates by reduced-pressure chemical vapor deposition, as reported elsewhere.[33] We first performed transport characterization of Hall bars (see the Supporting Information and Ref.[34]), the results of which attest to the high quality of the material and suggest that it is favorable for hosting hole quantum dots. The effective mass of holes in this system is found to be $m^* \sim 0.08\ m_0$ (where $m_0$ is the bare electron mass), close to that of electrons in GaAs ($m^* \sim 0.067\ m_0$). This implies that the spatial extent of the hole wavefunction will be wider than that of electrons in silicon, with an effective mass of $m_{Si} \sim 0.19\ m_0$. A relatively large spatial wavefunction could enhance tunnel couplings in quantum dots, facilitating qubit-qubit interaction, as well as relax lithographic constraints on gate geometries. In the following sections, we take the initial steps to explore these possibilities by demonstrating enhancement-mode single and double quantum dots in this material.

For the quantum dot studies, two devices—labelled Q1 and Q2— were prepared with Ohmic contacts and metallic top gate electrodes using a similar process as for the Hall bars. Device Q1 featured a SiGe buffer layer with 15% Si and was prepared with a gate oxide layer of 100 nm $Al_2O_3$ grown by atomic layer deposition, while device Q2 had an upper SiGe layer composed of 28% Si and a thinner gate oxide layer (24 nm $Al_2O_3$ + 1 nm $HfO_2$). The two samples were otherwise nominally identical. Both quantum dot samples were fabricated from wafers supporting peak mobility values in excess of $10^5$ cm$^2$/Vs. Electron-beam lithography was used to define a single-layer set of nanoscale gate electrodes that allow local accumulation of a two-dimensional hole gas (2DHG) and lateral confinement of holes by depletion to form quantum dots. The device design is similar to one employed previously in Si/SiGe quantum dot studies.[32,35]

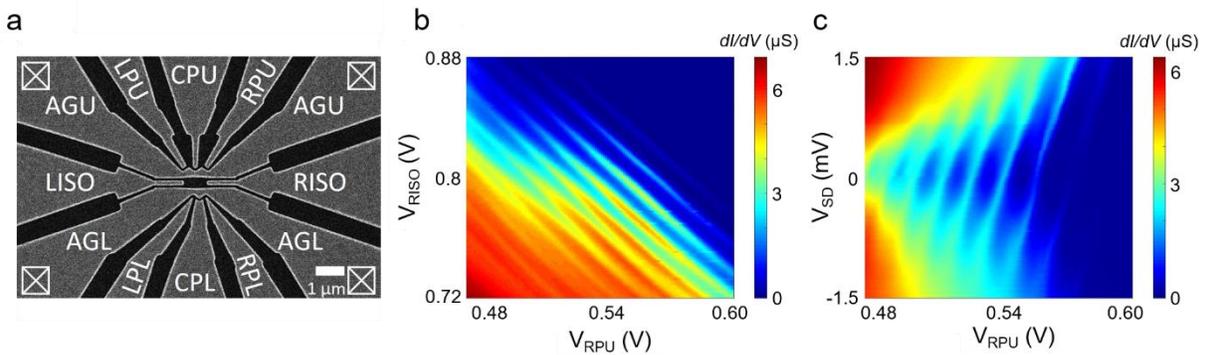

**Figure 1:** (a) Scanning electron micrograph of device Q1. Light gray areas are the Ti/Au gate electrodes, while the dark regions are the bare oxide. (b) Conductance map as a function of right isolation gate voltage ($V_{RISO}$) and upper right plunger gate voltage ($V_{RPU}$). Clear oscillations in the dot current are consistent with Coulomb blockade behavior, with the dot being emptied of holes one at a time as either gate voltage is increased. The separation between oscillation peaks can be used to extract the capacitance of each gate to the dot. (c) Dot conductance (color scale) plotted against source-drain bias ($V_{SD}$) and upper right plunger bias ($V_{RPU}$). Coulomb diamonds are clearly visible. Holes are removed from the quantum dot with increasing plunger gate voltage.

A scanning electron microscopy (SEM) image of sample Q1 is shown in Fig. 1(a). The wire-shaped upper and lower accumulation gates (AGU and AGL) accumulate holes in the quantum well layer, forming the two-dimensional hole gas (2DHG), and Ohmic contacts (shown as boxed X marks) are used to monitor the conductance through the wire or quantum dot. Plunger gates (upper and lower sets of left, center, and right plungers) allow tuning of the quantum dot size and dot-reservoir tunnel barrier heights, while left and right isolation gates (LISO, RISO) ensure that the upper and lower wires can be isolated from one another, and also serve as additional tuning parameters. The upper half of the device is designed to host two quantum dots, while the lower half is designed for only one. The gate electrodes are defined in a simple single metal layer, requiring just one step of electron-beam lithography and allowing flexible alignment tolerances.

The quantum dot studies were conducted in a dilution refrigerator at $T \sim 30$ mK, with similar behavior observed in the two measured samples. The upper accumulation gate (AGU) was negatively biased to accumulate holes under the reservoir areas and narrow wire connecting them, as shown in Fig. 1(a)). Typical threshold voltages for these devices range between -0.4 V and -1.6 V. Plunger and isolation gates CPU, RPU, and RISO were tuned to positive voltages to confine a quantum dot under the right segment of the upper wire. At the same time, a small ac bias of 100 µV at $f = 1333$ Hz was applied to the upper Ohmic contacts for monitoring the transport through the quantum dot. Charge stability diagrams of dot current as a function of various pairs of gate potentials were collected to assess the quantum dot behavior. Fig. 1 (b) shows an example of such a plot, tracking the dot current as a function of the voltages on depletion gates RISO and RPU. Clear diagonal Coulomb oscillations appear as the depletion gates are ramped through their respective ranges, corresponding to individual holes being removed from the dot as the gate bias is increased. In Fig. 1 (c), Coulomb diamonds are observed when the source-drain bias is tuned as a function of the potential on plunger gate RPU, with features resulting from the alignment of the dot chemical potential with that of either the source or drain electrode. The diamonds widen along the $V_{SD}$ axis moving from left to right, indicating increasing charging energy. In gated quantum dots with *n*-type carriers, this has been interpreted as evidence of reaching the low-occupancy regime.[36–38]

The experimental capacitance values between the dot and the individual gate electrodes can be extracted from the conductance maps. Here, the capacitance is $C = Q/V$, where $Q = +e$ is the elementary charge of one hole and $V$ is the extracted Coulomb oscillation period in volts along one gate's voltage axis. This procedure yields the capacitance values of each individual gate to the quantum dot. From the experimentally determined total capacitance, $C_{tot} = 92.8$ aF, a charging energy of $E_c = e^2/C_{tot} = 1.7$ meV is obtained, reasonably consistent with the height of the Coulomb diamonds in Fig. 1(c).

In addition to forming a single quantum dot, the device structure can also be configured to confine holes in two adjacent quantum dots, as lithographically defined in the upper device channel. Fig. 3 shows a stability diagram, collected using sample Q2, of dot conductance (color scale) as a function of upper left and upper right plunger gate potentials. The honeycomb-shaped anti-crossing features are consistent with the interplay between holes in the two separate quantum dots. We note that the potential landscape imposed by the gates naturally forms a double dot, and no bias is required on the center plunger gate, CPU, to achieve this configuration.

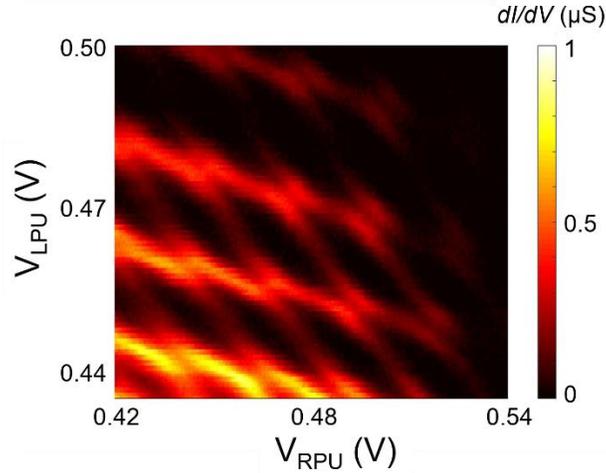

**Figure 2:** Stability diagram of right-side dot differential conductance (color scale) as a function of upper left plunger (LPU) and upper right plunger (RPU) gate biases. The honeycomb shaped anti-crossing features are consistent with coupling between two individual quantum dots.

The one-layer metal design allows for straightforward device fabrication and readily forms single and double quantum dots. Characteristic features of transport data from Ge/SiGe devices with this gate layout are broad Coulomb blockade transitions (Fig. 1 (b-c)) and the presence of only a handful (~10) of visible Coulomb oscillations between the featureless high-current regime (lower left of Fig. 1b) and the regime where transport is no longer visible (upper right of Fig. 1b). The broad features are not due to experimental factors such as voltage bias or temperature, but rather the broadening likely originates from highly transparent tunnel barriers created by the single metal layer design. To provide further insight into the origin of these features, we have performed a finite-element simulation of charge density using the geometry of sample Q1, based on the lateral dimensions extracted from the SEM image and the material stack thicknesses and dielectric properties. We model accumulation in the dot using self-consistent Thomas-Fermi screening and perform the calculation with COMSOL MULTIPHYSICS.[39,40] To tune up the dot in COMSOL, we start with the experimentally determined gate voltages and adjust them until we form a dot. Note that the model contains a threshold voltage at which the two-dimensional hole gas (2DHG) starts to accumulate, and we adjust this threshold voltage along with the gate voltages. The gate voltages for the tuned-up dot in COMSOL are all very similar to the experimental gate voltages in for the data in Fig. 1b, with the exception of the CPU gate, which likely indicates the presence of trapped charge near this gate in the actual device.

An example of the simulated charge density for the single dot case with ~1.5e of charge accumulated in the quantum dot is shown in Fig. 3. Far away from the quantum dot, the charge density is that of a 2DHG, as expected. On the right side of the expected confinement area, an oblong island of charge, *i.e.* a quantum dot, is present and appears more closely coupled to the right side 2DHG. Additional tuning of the various gate voltages within COMSOL indicates a rapid transition from no charge accumulation in the dot location to a flooded situation lacking charge confinement (not shown). The simulations indicate that confinement is only possible when a relatively small number of holes are present in the quantum dot, suggesting a shallow potential well relative to the tunnel barriers. The rapid transition from no charge accumulation in the quantum dot region to lack of charge confinement is qualitatively consistent with the characteristic experimental features seen in Fig. 1(b). To further explore the charge distribution in COMSOL, we determine the capacitances between the dot and the gates by varying the gate voltage slightly and measuring the corresponding change in dot charge. The results are shown in Fig. 3(c), along with the experimentally determined capacitance values. The COMSOL simulated capacitance values are in reasonable agreement with the experimental results, suggesting a lithographic quantum dot in the experiment that has similar lateral extent as seen in the simulation.[41]

The qualitative agreement between the simulations and the experimental observations indicates that the gate layout on this material stack produces a quantum confinement potential which is broad and shallow with very transparent tunnel barriers. As suggested by the simulations, such a potential is only able to confine a small number of charge carriers. This is consistent with the experimental observations in Fig. 1(c) where the Coulomb diamonds grow in size as the number of holes in the quantum dot is reduced, which is commonly interpreted as the low-occupancy regime in laterally confined semiconductor quantum dots.[36–38,42] We note that the tunnel-broadening excludes the possibility of observing excited states in this structure.

Finally we note, the gate layouts employed here were inspired by those successfully used in Si/SiGe for electron quantum dots,[32,35] and the fact that a similar device design can be utilized in both Si/SiGe and Ge/SiGe opens the possibility for direct comparisons between the systems of electrons in silicon and holes in germanium. For example, experiments investigating tunnel couplings in these two systems will elucidate the contribution of the effective mass (shown here to be ~2.4 × lighter in the hole system) on enhancing tunneling coupling in quantum dot systems. The complementary systems of Si/SiGe and Ge/SiGe provide a new avenue for exploring quantum dot properties relevant to spin qubits.

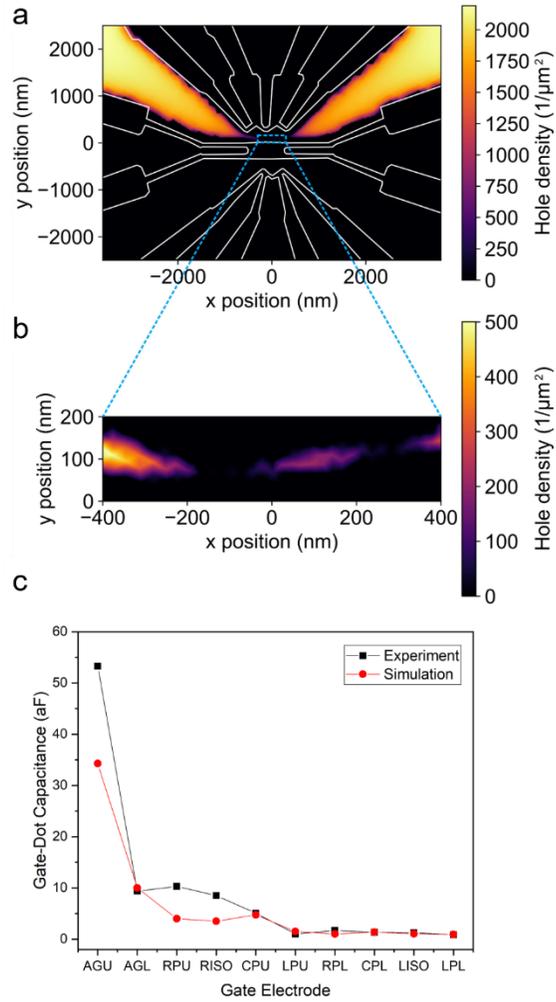

**Figure 3:** Calculated charge density map based on the single quantum dot configuration in Fig. 2. Panel (a) shows the gate electrodes outlined in white, with charge primarily concentrated under the left and right sides of the upper accumulation gate and a small quantum dot formed between these reservoirs. Panel (b) shows a zoomed-in view of the oblong charge distribution of the quantum dot (between approximately 0 – 200 nm in the x-direction), with the tips of the reservoirs visible at the left and right edges of the field of view. Panel (c) shows a comparison of gate-dot capacitance values obtained from experiment (black squares) and simulation (red circles).

CONCLUSIONS

We have demonstrated lithographically defined single and double hole quantum dots in high quality strained Ge/SiGe quantum wells. These results strongly suggest that this material system may serve as a viable host for spin-based qubits (compatible with CMOS processing) and enable quantitative comparisons between quantum dots in electron and hole systems. Multi-metal-layer devices, as are commonly used in Si/SiGe, should be directly applicable to Ge/SiGe to increase the sharpness of tunnel barriers and provide more orthogonal control of coupling between adjacent quantum dots. Development of successful qubit architectures will ultimately call for in-depth studies of qubit decoherence mechanisms in this system, in particular the impact of charge noise due to the enhanced spin-orbit coupling.

AUTHOR INFORMATION

**Corresponding Author**

*E-mail (W.J. Hardy): wilhard@sandia.gov

**Notes**

The authors declare no competing financial interest.

ACKNOWLEDGMENTS

This work was funded, in part, by the Laboratory Directed Research and Development Program and performed, in part, at the Center for Integrated Nanotechnologies, an Office of Science User Facility operated for the U.S. Department of Energy (DOE) Office of Science. Sandia National Laboratories is a multi-mission laboratory managed and operated by National Technology &

Engineering Solutions of Sandia, LLC, a wholly owned subsidiary of Honeywell International Inc., for the U.S. Department of Energy's National Nuclear Security Administration under contract DE-NA-0003525. The views expressed in this article do not necessarily represent the views of the U.S. Department of Energy or the United States Government. The work at NTU was supported by the Ministry of Science and Technology (107-2622-8-002-018-).

Supporting Information for

# Single and Double Hole Quantum Dots in Strained Ge/SiGe Quantum Wells

*Will J. Hardy[a]\*, C. T. Harris[a], Y.-H. Su[b], Y. Chuang[b], J. Moussa[a], Leon Maurer[a], Jiun-Yun Li[b], Tzu-Ming Lu[a], and Dwight R. Luhman[a].*

[a]Sandia National Laboratories, Albuquerque, NM 87123, United States

[b]Department of Electrical Engineering, National Taiwan University, Taipei, Taiwan

**I. Material Growth Details**

The material stack consists of a Si buffer layer (200 nm) grown on the bare 8-inch diameter p-type Si (001) substrate, low-temperature Ge virtual substrate (100 nm), high-temperature relaxed Ge buffer (200 nm), lower $Si_xGe_{1-x}$ graded buffer, strained Ge quantum well layer (20 nm), and upper $Si_xGe_{1-x}$ barrier (70 nm). Further details can be found in Ref.[SR1].

**II. Effective Mass Characterization**

Two Hall bar devices (labelled H1 and H2) were prepared from wafers with $x = 0.28$ and $x = 0.36$ using standard microfabrication processes.[SR2] Transport measurements of zero-field longitudinal and Hall resistances were performed in a pumped $^4$He cryostat to assess the material quality. Peak mobility values of $\mu \gtrsim 6 \times 10^4$ cm$^2$/Vs were measured separately in a $^3$He cryostat at T = 0.3 K. From the mobility measurements, we extract the carrier mean free path as a function of carrier density, shown in Fig. S1 for the same wafers as used in (a) sample Q1; (b) samples H1 and Q2; and (c) sample H2.

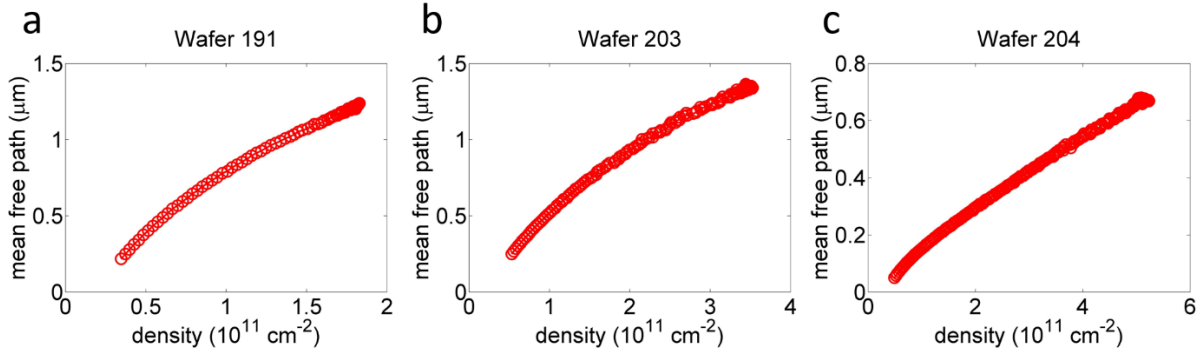

**Figure S1:** Mean free path as a function of carrier density measured using the same wafers as for (a) sample Q1; (b) samples H1 and Q2; and (c) sample H2.

Magnetoresistance data at low temperatures reveal Shubnikov-de Haas (SdH) oscillations that grow in amplitude as the temperature is decreased, with the measurement repeated at several values of carrier density by varying the top gate voltage. The observation of SdH oscillations immediately confirms the low disorder of the material, and quantitative analysis allows extraction of effective mass information. First, the magnetic field values are converted to filling factor ν, and a slowly varying classical MR background is subtracted from the raw data, as shown for an example value of carrier density $p = 3.66 \times 10^{11}$ cm$^{-2}$ in Fig. S1(a). We then consider a window of magnetic field in which the field is large enough for the oscillations to be clear, yet small enough to avoid Zeeman splitting of the peaks. The amplitudes of the even filling factor dips in the resistance oscillations are extracted, and the evolution with temperature is traced over a range of carrier density values, the latter having been determined from the Hall resistance slopes. The temperature dependence of the oscillation amplitude is fit using the following Lifshitz-Kosevich expression:[SR2,SR3]

$$\Delta R_{xx} \propto \frac{2\pi^2 \frac{k_B T}{\Delta E}}{\sinh\left[2\pi^2 \frac{k_B T}{\Delta E}\right]},$$

where $\Delta E = heB/2\pi m^*$, $h$ is Planck's constant, and $k_B$ is Boltzmann's constant. This yields the effective mass at each density value, as shown in Fig. S1(b) along with an example of the fit to the temperature dependence of the oscillation amplitude (inset, $p = 3.67 \times 10^{11}$ cm$^{-2}$, $\nu = 18$). The hole carriers in this system have an effective mass $m^* \sim 0.08\ m_0$, where $m_0$ is the free electron mass, independent of density over the carrier density range $p = 2 - 5 \times 10^{11}$ cm$^{-2}$. This is consistent with an initial report in this system that focused on higher densities,[SR4] and is also close to a theoretically predicted value of $m^* \sim 0.05\ m_0$.[SR5]

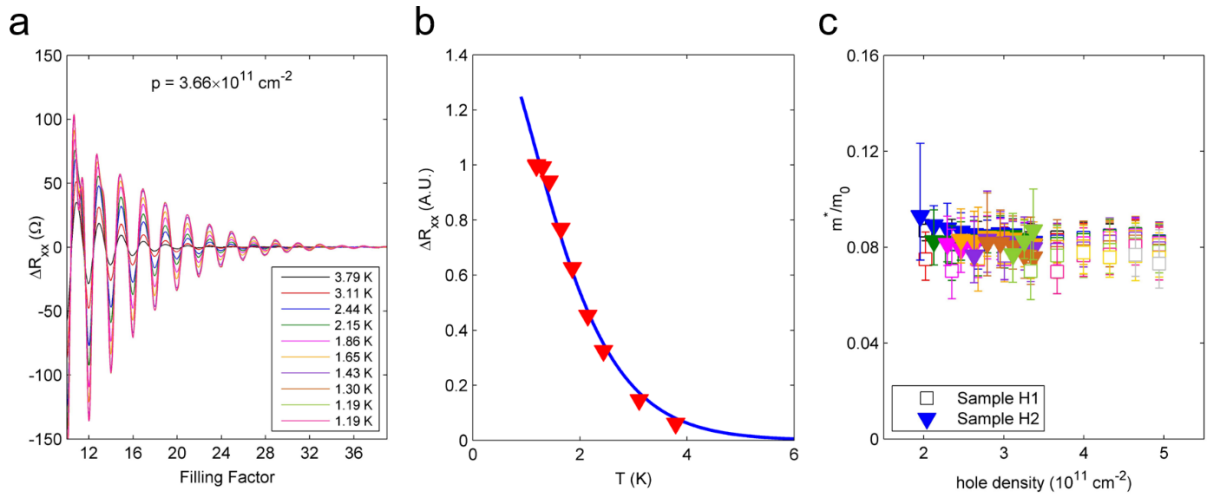

**Figure S2:** (a) Example of Shubnikov-de Haas (SdH) oscillations at selected temperatures, measured at a fixed hole density of $p = 3.66 \times 10^{11}$ cm$^{-2}$ in sample H1. The horizontal axis scale has been converted from units of magnetic field to filling factor, and a slowly varying background has been subtracted. (b) Example of the SdH oscillation amplitude versus temperature data (red triangles) and fit (blue curve) used to obtain the effective mass values (example data measured at $p = 3.67 \times 10^{11}$ cm$^{-2}$, $\nu = 18$). (c) Scaled hole effective mass $m^*/m_0$ as a function of hole density, as observed in samples H1 (open squares) and H2 (filled triangles). No clear dependence of effective mass on filling factor $\nu$ was observed within the scatter of the data (various studied values of $\nu$, in the range of $\nu = 12 - 34$, represented by the colors of the data points). The effective mass $m^* \sim 0.08 m_0$ is nearly constant over the measured range of densities, with good quantitative agreement between the two samples.